\documentclass[preprintnumbers,amsmath,amssymb,showpacs]{revtex4}

\usepackage{graphicx}
\usepackage{dcolumn}
\usepackage{bm}
\usepackage{graphicx}
\usepackage{booktabs}
\usepackage{color} 

\begin{document}

\title{Efficient scheme for three-photon Greenberger-Horne-Zeilinger state generation}
\author{Dong Ding$^{1,2}$}
\author{Fengli Yan$^1$}
 \email{flyan@hebtu.edu.cn}

\affiliation {$^1$ College of Physics Science and Information Engineering, Hebei Normal University, Shijiazhuang 050024, China\\
$^2$Department of Basic Curriculum, North China Institute of Science and Technology, Beijing 101601, China}

\date{\today}

\begin{abstract}
We propose an efficient scheme for the generation of three-photon Greenberger-Horne-Zeilinger (GHZ) state with linear optics and postselection. Several devices are designed and a two-mode quantum nondemolition (QND) detection is introduced to obtain the desired state. It is worth noting that the states which have entanglement in both polarization and spatial degrees of freedom are created in one of the designed setups. The method described in the present scheme can create a large number of three-photon GHZ states in principle. We also discuss an approach to generate the desired GHZ state in the presence of channel noise.
\end{abstract}


\maketitle

\section{Introduction}

Entanglement is considered to be the most nonclassical manifestation of quantum formalism that has yet been put into use as a new resource and is as real as energy \cite{Quantum entanglement RevModPhys}. Since the seminal work by Einstein, Podolsky, and Rosen \cite{EPR1935} there
has been extensive research on entanglement. In 1991, Ekert proposed a quantum cryptography scheme based on Bell's theorem \cite{Ekert91}, which was the first discovery for quantum
information theory that involved entanglement. There are many other interesting applications that incorporate entanglement in quantum information theory, such as quantum teleportation \cite{B93,
Boschi}, quantum key distribution \cite{B92}, quantum secure direct communication \cite{SI1999,dengLong2003}, etc. So far, the most used source of entanglement is entangled-photon states, which are generated by a nonlinear process of parametric down-conversion (PDC) source \cite{PDC1995}.

Although there exist several types of multipartite entangled states, for example the Greenberger-Horne-Zeilinger (GHZ) states \cite{GHZ1990}, W states \cite{W2000} and cluster states \cite{cluster2001}, we will restrict our discussion to the facet of three-photon GHZ states.
A maximally entangled GHZ state shows perfect correlation properties for several observing parties. Using the  GHZ
states, one can achieve quantum communication and quantum computation schemes \cite{CB1997,GYJPA05,GYEPRA}. GHZ entanglement also plays an important role in fundamental tests of quantum mechanics versus local realism. Bouwmeester \emph{et al}. \cite{GHZExperiment99}  presented experimental evidence for the observation of three-photon GHZ entanglement. Using two pairs of entangled photons, three-photon GHZ state
\begin{equation}\label{GHZ state}
{\left| {{\phi ^ + }} \right\rangle _{ABC}} =
\frac{1} {{\sqrt 2 }}{\left( {\left| {HHV} \right\rangle + \left|
{VVH} \right\rangle } \right)_{ABC}}
\end{equation}
has been observed experimentally, where $H$ ($V$) denotes horizontal (vertical) polarization and the subscripts $A$, $B$ and $C$ respectively represent three photons sent to the three observing locations. In their experiment, the GHZ entanglement is observed under the condition that both the trigger photon and the three entangled photons are actually detected (simultaneously).

Additionally, there have been studies of nondestructive quantum nondemolition (QND) \cite{Milburn1984, Imoto1985, QND2004}. In 2004, Nemoto and Munro \cite{QND2004} proposed a scheme for constructing a near deterministic controlled-NOT (CNOT) gate using the QND detection, which can evolve the combined system of signal and probe modes and functions as both a single-photon detector and a two-qubit polarization parity detector but does not destroy the signal
photons. The heart of the QND detection are weak cross-Kerr nonlinearities that may be generated by electromagnetically induced transparency \cite{CKavailable1996}. The cross-Kerr nonlinearity has
a Hamiltonian of the form \({\mathcal{H}_{QND}} = \hbar \chi
a_s^\dag {a_s}a_p^\dag {a_p}\),  where $a_s^\dag$ and ${a_s}$
represent the creation and annihilation operators of the signal
mode, while $a_p^\dag$ and ${a_p}$ stand for the creation and
annihilation operators of the probe mode, and $\chi$ is the coupling
strength of the nonlinearity. Assuming a signal state \({\left| \psi
\right\rangle _s} = {c_0}{\left| 0 \right\rangle _s} + {c_1}{\left|
1 \right\rangle _s}\) and an initially coherent probe beam $|\alpha\rangle_p$,
 the combined system evolves as
 \({\texttt{e}^{\texttt{i}{\mathcal{H}_{QND}}t/\hbar }}{\left| \psi  \right\rangle _s}{\left| \alpha  \right\rangle _p}
 = {c_0}{\left| 0 \right\rangle _s}{\left| \alpha  \right\rangle _p} + {c_1}{\left| 1 \right\rangle _s}{\left|
 {\alpha {\texttt{e}^{\texttt{i}\theta }}} \right\rangle _p}\), where \(\theta  = \chi t\) and $t$ is the
 interaction time. Generally speaking, the probe beam $|\alpha\rangle_p$ accumulates
 a phase shift $\left| {\alpha {\texttt{e}^{n\texttt{i}\theta }}} \right\rangle_p$ directly proportional to the number of photons $n$ in the signal mode. One can immediately read out, but not destroy the signal state in the process, the phase shift using an $X$ homodyne measurement.

Although entanglement plays a central role in quantum information processing, purely entangled states may become degraded when they are transmitted through a noisy channel. Before the states are used to communicate reliably, entanglement purification must be done to obtain almost perfectly entangled states from the mixed ones. In 1996, Bennett \emph{et al}. \cite{EEP96} presented an entanglement purification protocol (EPP) which is used to purify a Werner state based on CNOT gates. Several improved schemes of both single-pair and multipartite quantum systems were subsequently proposed \cite{DeutschEEP96,
MuraoEEP1998}. In these schemes, one can increase the entanglement and improve fidelity of quantum states by repeatedly performing the purification protocol. In 2002, Simon and Pan \cite{SimonPan2002} proposed an EPP for two entangled photons with two PDC sources and two polarizing beam splitters. Recently,
several improved schemes were proposed \cite{Lixh2010,ShengDeng2010}, where two one-step deterministic
entanglement purification protocols have been presented with simple linear optical elements. More recently, Deng \cite{dengfuguo2011} proposed an error correction protocol for multipartite polarization entanglement using spatial entanglement. However, the creation of GHZ entanglement in both polarization and spatial degrees of freedom has not been reported.

In this paper, we first report a new scheme for the creation of two three-photon states which are entangled in both polarization and spatial degrees of freedom based on PDC sources and nondestructive QND detection. The present setup is composed of five polarizing beam splitters, two PDC sources consisting of a $\beta$-barium-borate (BBO) crystal and a mirror, and a QND detector. Then, an improved device can immediately transform any one of the created states into a desired three-photon GHZ state. Furthermore, in the presence of channel noise the design is still valid.

\section{Creation of polarization entanglement and spatial entanglement }
Before we describe the proposed scheme, let us first explain the creation of polarization entanglement with a PDC source. The core element of the PDC source is a nonlinear crystal.  Pairs of polarization entangled photons \begin{equation}\label{}
\frac{1}
{{\sqrt 2 }}\left( {{{\left| H \right\rangle }_a}{{\left| V \right\rangle }_b} - {{\left|
 V \right\rangle }_a}{{\left| H \right\rangle }_b}} \right)
\end{equation}
in modes $a$ and $b$ are generated with a certain probability by a short pulse of ultraviolet light which passes through the crystal \cite{PDC1995}.

\begin{figure}
  \includegraphics[width=5in]{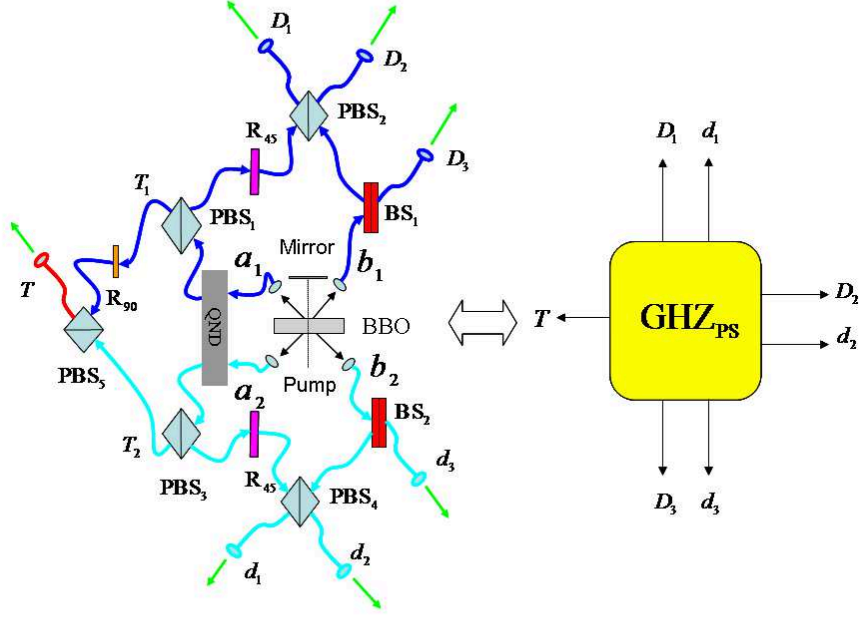}\\
  \caption{(color online). Schematic drawing of the creation of polarization entanglement and spatial entanglement. Two parametric down-conversion sources (a BBO crystal and a mirror) are used to produce two pairs of photons that have both polarization and spatial entanglement. A nondestructive quantum nondemolition (QND) detector is used to discriminate between the two states (\ref{up down}) and (\ref{four-mode}). We simplify the device by labeling it as {\sc GHZps}.}\label{ghzps}
\end{figure}

We now investigate only the cases where two pump photons pass through the BBO crystal to yield two entangled photon pairs. This is represented by the product state
 \begin{equation}\label{product state}
\frac{1}
{2}\left( {{{\left| H \right\rangle }_{a_i}}{{\left| V \right\rangle }_{b_i}} - {{\left| V \right\rangle }_{a_i}}
{{\left| H \right\rangle }_{b_i}}} \right)\left( {{{\left| H \right\rangle }_{a_j}}{{\left| V \right\rangle }_{b_j}}
- {{\left| V \right\rangle }_{a_j}}{{\left| H \right\rangle }_{b_j}}} \right),
\end{equation}
 where the subscripts $a_{i,j}$ and $b_{i,j}$ ($i,j=1,2$) respectively represent four spatial modes, say the upper modes (produced by a pump pulse coming from below and traversing BBO crystal) $a_1$ and $b_1$ and the lower modes (produced by a pump pulse reflected by a mirror and traversing BBO crystal a second time) $a_2$ and $b_2$. As shown in Fig.1, there is a mirror in our setup. Because of the mirror, there must be three cases. Case 1, for $i=j=1$, the two entangled photon pairs are in the upper modes $a_1$ and $b_1$.
Case 2, for $i=j=2$, the two entangled photon pairs are in the lower modes $a_2$ and
$b_2$. Case 3, for $i \neq j$, one entangled photon pair is in the upper modes $a_1$ and $b_1$ and the other one is in the lower modes $a_2$ and $b_2$. The fiber-based circuit is sketched in Fig.1. Each polarizing beam splitter (PBS) is used to transmit the $|H\rangle$ polarization photons and reflect the $|V\rangle$ polarization photons. Because any information as to which pair each photon belongs to is erased at the PBS, it is impossible in principle to determine whether a photon has been transmitted or reflected since one obtains the photon after the PBS. The beam splitter (BS) is a $50:50$ polarization-independent beam splitter. When a photon travels to the BS, it has a 50\% chance of being transmitted or reflected. The half wave plate $ \texttt {R}_{45}$  represents a Hadamard operation and it is used to transform the polarization states $|H\rangle$ and $|V\rangle$  into $\frac{1}
{{\sqrt 2 }}\left( {{{\left| H \right\rangle }} + {\left| V \right\rangle }} \right)$ and $\frac{1}
{{\sqrt 2 }}\left( {{{\left| H \right\rangle }} - {\left| V \right\rangle }} \right)$, respectively.
The half wave plate $ \texttt {R}_{90}$ can convert the polarization state $|V\rangle$ into $|H\rangle$ or vice versa.
Through the apparatus towards the spatial modes $T$, $D_i$, and $d_i$ ($i=1,2,3$) the individual components of Eq.(\ref{product state}) evolve as
 \begin{equation}\label{}
|H\rangle_{a_{1}} \rightarrow |H\rangle_{T_{1}} \equiv |T\rangle,~~~
|H\rangle_{a_{2}} \rightarrow |H\rangle_{T_{2}} \equiv |T\rangle,
\end{equation}
 \begin{equation}\label{}
|V\rangle_{a_{1}} \rightarrow \frac{1} {{\sqrt 2 }}\left(\left| V \right\rangle_{D_{1}} + \left| H \right\rangle_{D_{2}}\right), ~~~
|V\rangle_{a_{2}} \rightarrow \frac{1} {{\sqrt 2 }}\left(\left| V \right\rangle_{d_{1}} + \left| H \right\rangle_{d_{2}}\right),
\end{equation}
\begin{equation}\label{}
|H\rangle_{b_{1}} \rightarrow \frac{1} {{\sqrt 2 }}\left(\left| H \right\rangle_{D_{1}} + \left| H \right\rangle_{D_{3}}\right),~~~
|H\rangle_{b_{2}} \rightarrow \frac{1} {{\sqrt 2 }}\left(\left| H \right\rangle_{d_{1}} + \left| H \right\rangle_{d_{3}}\right),
\end{equation}
\begin{equation}\label{}
|V\rangle_{b_{1}} \rightarrow \frac{1} {{\sqrt 2 }}\left(\left| V \right\rangle_{D_{2}} + \left| V \right\rangle_{D_{3}}\right),~~~
|V\rangle_{b_{2}} \rightarrow \frac{1} {{\sqrt 2 }}\left(\left| V \right\rangle_{d_{2}} + \left| V \right\rangle_{d_{3}}\right),
\end{equation}
where $|T\rangle$ represents trigger photon, which may come from $\left| H \right\rangle _{T_{1}}$ or $\left| H\right\rangle _{T_{2}}$.

Now we discuss the creation of polarization entanglement and spatial entanglement using the equipment shown in Fig.1. For $i=j$, the state (\ref{product state}) evolves as
\begin{equation}\label{up down}
\frac{1}
{2}{\left| T \right\rangle}\left( {{{\left| H \right\rangle }_{{D_1}}}
{{\left| H \right\rangle }_{{D_2}}}{{\left| V \right\rangle }_{{D_3}}} +
{{\left| V \right\rangle }_{{D_1}}}{{\left| V \right\rangle }_{{D_2}}}
{{\left| H \right\rangle }_{{D_3}}} + {{\left| H \right\rangle }_{{d_1}}}
{{\left| H \right\rangle }_{{d_2}}}{{\left| V \right\rangle }_{{d_3}}} +
{{\left| V \right\rangle }_{{d_1}}}{{\left| V \right\rangle }_{{d_2}}}
{{\left| H \right\rangle }_{{d_3}}}} \right)
.\end{equation}
While for $i \neq j$, the state (\ref{product state}) evolves as
\begin{equation}\label{four-mode}
\frac{1} {2}{\left| T \right\rangle}\left( {{{\left| H
\right\rangle }_{{d_1}}}{{\left| H \right\rangle }_{{d_2}}}{{\left|
V \right\rangle }_{{D_3}}} + {{\left| V \right\rangle
}_{{d_1}}}{{\left| V \right\rangle }_{{D_2}}}{{\left| H
\right\rangle }_{{d_3}}} + {{\left| H \right\rangle
}_{{D_1}}}{{\left| H \right\rangle }_{{D_2}}}{{\left| V
\right\rangle }_{{d_3}}} + {{\left| V \right\rangle
}_{{D_1}}}{{\left| V \right\rangle }_{{d_2}}}{{\left| H
\right\rangle }_{{D_3}}}} \right) .\end{equation}
The $\left| T \right\rangle$ components in Eqs.(\ref{up down}) and (\ref{four-mode}) will be ignored for clarity in the following formal derivation (or presentation). The three-photon entangled state described in Eq.(\ref{up down}) contains entanglement in both polarization and spatial degrees of freedom, and it is the superposition of the states
$\frac{1}{{\sqrt 2 }}\left( {{{\left| H \right\rangle
}_{{D_1}}}{{\left| H \right\rangle }_{{D_2}}}{{\left| V
\right\rangle }_{{D_3}}} + {{\left| V \right\rangle
}_{{D_1}}}{{\left| V \right\rangle }_{{D_2}}}{{\left| H
\right\rangle }_{{D_3}}}} \right) $ and $\frac{1}{{\sqrt 2 }}\left(
{{{\left| H \right\rangle }_{{d_1}}}{{\left| H \right\rangle
}_{{d_2}}}{{\left| V \right\rangle }_{{d_3}}} + {{\left| V
\right\rangle }_{{d_1}}}{{\left| V \right\rangle }_{{d_2}}}{{\left|
H \right\rangle }_{{d_3}}}} \right)$. Here  $D_1$, $D_2$, $D_3$ and
$d_1$, $d_2$, $d_3$ denote different spatial modes. Obviously, this state is a product state with two degrees of freedom and can be written as
\begin{equation}\label{density matrix}
\rho=\rho_{P}\otimes\rho_{S},
\end{equation}
where $\rho_{P}$ and $\rho_{S}$ are reduced density operators and respectively represent three-photon entangled states in the polarization degree of freedom and the spatial degree of freedom. While the other three-photon state shown in Eq.(\ref{four-mode}) is the superposition of the state $ \frac {1}{\sqrt
2}({{{\left| H \right\rangle }_{{d_1}}}{{\left| H \right\rangle
}_{{d_2}}}{{\left| V \right\rangle }_{{D_3}}} + {{\left| V
\right\rangle }_{{d_1}}}{{\left| V \right\rangle }_{{D_2}}}{{\left|
H \right\rangle }_{{d_3}}} } )$ (two photons passing through the
lower modes and one photon going through the upper mode) and the
state $\frac {1}{\sqrt 2}({{{\left| H \right\rangle
}_{{D_1}}}{{\left| H \right\rangle }_{{D_2}}}{{\left| V
\right\rangle }_{{d_3}}} + {{\left| V \right\rangle
}_{{D_1}}}{{\left| V \right\rangle }_{{d_2}}}{{\left| H
\right\rangle }_{{D_3}}}})$ (two photons passing through the upper
modes and one photon going through the lower mode).

We should note that the state shown in Eq.(\ref{four-mode}) has not been mentioned in the previous. This state has entanglement in both the polarization and the spatial degrees of freedom, while it cannot be written as the presentation shown in Eq.(\ref{density matrix}). The entanglement resources described in Eqs.(\ref{up down}) and (\ref{four-mode}) may be used for some practical applications as was previously stated. In this scheme we can obtain the desired three-photon GHZ state based on the two entanglements, while they need to be distinguished in a nondestructive way.

In order to distinguish the state (\ref{up down}) connecting with the cases 1
and 2 from the state (\ref{four-mode}) connecting with case 3, we use the QND detection as shown in Figs.1 and
2. The main idea is that one can decide whether there is one photon in each
spatial mode $a_1$, $b_1$, $a_2$ and $b_2$ or not by deciding whether there is one
photon in each spatial mode $a_1$ and $a_2$ or not before $\texttt{PBS}_1$ and $\texttt{PBS}_3$. Obviously, under the condition that both the trigger photon and the three entangled photons are detected
accurately by a fourfold coincidence detector, the two photons going through the two modes $a_1$ and $a_2$, or only the spatial mode $a_1$ ($a_2$),
must be in a superposition of the states $|HV\rangle_{a_1a_1}$, $|HV\rangle_{a_2a_2}$,
$|HV\rangle_{a_1a_2}$, and $|VH\rangle_{a_1a_2}$.
However, in order to remain in spatial entanglement we
never distinguish the state $|HV\rangle_{a_1a_1}$ from $|HV\rangle_{a_2a_2}$,
or $|HV\rangle_{a_1a_2}$ from $|VH\rangle_{a_1a_2}$,
otherwise, the spatial entanglement would be destroyed. That is, we
only need to distinguish $|HV\rangle_{a_1a_1}$ and
$|HV\rangle_{a_2a_2}$ from  $|HV\rangle_{a_1a_2}$ and
$|VH\rangle_{a_1a_2}$.  This is the key trick of the scheme, and
the task can be accomplished simply by using cross-Kerr nonlinearities.

As shown in Fig.2, the QND detector is composed of four PBSs and two
cross-Kerr nonlinearities. After all these
interactions, there are no net phase shifts for the
terms of $|HV\rangle_{a_1a_2}$ and $|VH\rangle_{a_1a_2}$, while
there is a phase shift $\theta$ or $-\theta$ for $|HV\rangle_{a_1a_1}$
or $|VH\rangle_{a_2a_2}$. In terms of the Eq.(2) in Ref.\cite{QND2004}, with an $X$ homodyne
measurement, one can distinguish the superposition of the
states $|HV\rangle_{a_1a_2}$ and $|VH\rangle_{a_1a_2}$ from that of
$|HV\rangle_{a_1a_1}$ and $|HV\rangle_{a_2a_2}$, but the states
$|HV\rangle_{a_1a_1}$ and $|HV\rangle_{a_2a_2}$, or $|HV\rangle_{a_1a_2}$ and $|VH\rangle_{a_1a_2}$, cannot be distinguished. Clearly, let $X_0$ represents the midpoint between two peaks of probability amplitudes associated with the outputs of the homodyne measurement. Then, for \(X < {X_0}\), the two pairs emit each in the upper modes $a_1$ and $b_1$ or each in the lower modes $a_2$ and $b_2$ corresponding to Eq.(\ref{up down}).
While for \(X > {X_0}\), there is one photon in each spatial mode $a_1$, $b_1$, $a_2$, and $b_2$ corresponding to Eq.(\ref{four-mode}).
In addition, for the case of \(X <{X_0}\), a phase shift operation $\phi(x)$ corresponding to $x$ (the value of the $X$ homodyne measurement) should be performed to erase the accompanying effect of phase shift. For more details of QND detection are described by Nemoto \emph{et al}. in Ref.\cite{QND2004}.

\begin{figure}
  \includegraphics[width=3.5in]{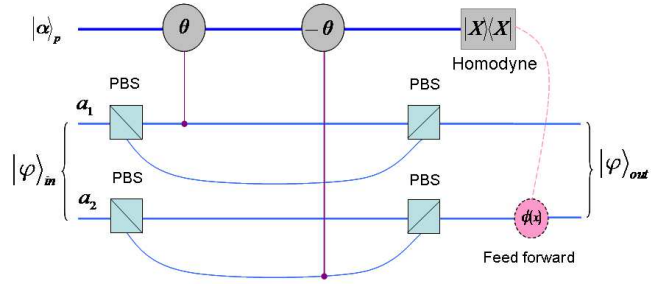}\\
  \caption{(color online). A quantum nondemolition detector. $a_1$ and $a_2$ are two spatial modes shown in Fig.1. For each spatial mode, a polarization encoded photon is split into one of the two paths at a polarizing beam splitter. After interactions with cross-Kerr nonlinearities, an $X$ homodyne measurement occurs on the probe beam $|\alpha\rangle_p$. The result is that two states (\ref{up down}) and (\ref{four-mode}) can be distinguished, while the two photons (signal modes) are nearly unaffected. A phase shift operation $\phi(x)$ dependent on the measurement result $x$ is performed via a feed forward process to eliminate the accompanying effect of phase shift.}\label{qnd}
\end{figure}

So far, we have considered the case where both the trigger photon and three
entangled photons are detected accurately (i.e., a four-mode coincidence detector clicks, for details see the next section). We have shown how to successfully generate and distinguish entangled states (\ref{up down}) and (\ref{four-mode}), which are in both the polarization entanglement and the spatial entanglement. In the next section, we will use these states to generate substantial three-photon GHZ states with, and without, channel noise.

\section{Generation of three-photon GHZ state}

As described above, we have shown that both the polarization entanglement and the spatial entanglement could be naturally produced in our device {\sc GHZps}. Because of these advantages, one may implement polarization entanglement purification using spatial entanglement \cite{SimonPan2002, Lixh2010, ShengDeng2010}. This part of the procedure will be relevant for the generation of three-photon GHZ state, so let us consider the setup shown in Fig.3, which has been inspired from the scheme of error correction presented by Deng in Ref.\cite{dengfuguo2011}. The general idea of our improved scheme is to construct an efficient apparatus to create a large number of three-photon GHZ states, even if there exists channel noise. The heart of this device is a {\sc GHZps}, which is used to provide initial states (both polarization and spatial entanglement). Three half wave plates $ \texttt {R}_{90}$s in spatial modes $D_1$, $D_2$, and $D_3$ and three PBSs are introduced to evolve the initial states. Finally, under the condition that the four-photon coincidence detector clicks (i.e. each single-photon detector registers a photon simultaneously), a three-photon GHZ state (\ref{GHZ state}) is created at a rate up to that predicted by local unitary transformations.

\begin{figure}
  \includegraphics[width=3in]{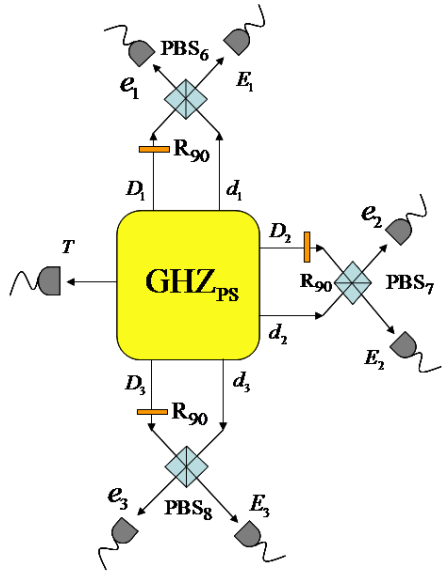}\\
  \caption{(color online). The schematic diagram of generation of three-photon GHZ state. The heart of this setup is a {\sc GHZps} shown in Fig.1. The functions of the PBSs and the half wave plates $ \texttt {R}_{90}$s are same as the former (see context).}\label{ghz}
\end{figure}

For the subsequent discussion, we first consider the case with one photon in each spatial mode $a_1$, $b_1$, $a_2$, and $b_2$. An initial state can be written as
\begin{equation}\label{four-mode2}
{\left| {{\psi ^ + }} \right\rangle _{PS}} = \frac{1}
{2}\left( {{{\left| {HHV} \right\rangle }_{{d_1}{d_2}{D_3}}} + {{\left| {HHV} \right\rangle }_{{D_1}{D_2}{d_3}}} + {{\left| {VVH} \right\rangle }_{{d_1}{D_2}{d_3}}} + {{\left| {VVH} \right\rangle }_{{D_1}{d_2}{D_3}}}} \right),
\end{equation}
where the subscripts $P$ and $S$  represent polarization and spatial modes. After three photons pass $\texttt{PBS}_6$, $\texttt{PBS}_7$ and $\texttt{PBS}_8$, respectively, the state $\left| {{\psi ^ + }} \right\rangle _{PS}$ evolves to
$\frac{1} {2}[{\left( {\left| {HHH} \right\rangle  + \left| {VVV}
\right\rangle } \right)_{{e_1}{e_2}{E_3}}} + {\left( {\left| {HVV}
\right\rangle  + \left| {VHH} \right\rangle }
\right)_{{E_1}{E_2}{e_3}}}]$, where $e_i$ and $E_i$ ($i=1,2,3$) denote different spatial modes corresponding to three photons $A$ ($i=1$), $B$ ($i=2$), and $C$ ($i=3$), respectively. Since each spatial mode is connected to a nondestructive single-photon detector, also we can use any one of the $e_i$ or $E_i$ to describe a nondestructive single-photon detector. These nondestructive single-photon detectors, as well as the trigger-photon detector $T$, are used to construct four-mode coincidence detectors.
Clearly, the state ${\left| {\phi _0^ + } \right\rangle _{ABC}} = \frac{1}
{{\sqrt 2 }}{\left( {\left| {HHH} \right\rangle + \left| {VVV}
\right\rangle } \right)_{ABC}}$ or the state ${\left| {{\phi_1^ + }} \right\rangle _{ABC}} =
\frac{1} {{\sqrt 2 }}{\left( {\left| {HVV} \right\rangle + \left|
{VHH} \right\rangle } \right)_{ABC}}$ will be obtained if the four
photons are all detected simultaneously, one by each detector $T$,
$e_1$, $e_2$, and $E_3$ or $T$, $E_1$, $E_2$, and $e_3$. That is, a four-photon coincidence detector acts as a projective measurement onto a GHZ state and filters out the undesirable terms. By
performing a bit-flip operation ${\sigma _x} = \left| H \right\rangle
\left\langle V \right| + \left| V \right\rangle \left\langle H
\right|$ on the photon $C$ ($B$), we can immediately obtain the desired GHZ entanglement ${\left| {{\phi ^ + }} \right\rangle _{ABC}} =
\frac{1} {{\sqrt 2 }}{\left( {\left| {HHV} \right\rangle + \left|
{VVH} \right\rangle } \right)_{ABC}}$ from the state ${\left| {\phi _0^ + }
\right\rangle _{ABC}}$ (${\left| {\phi _1^ + } \right\rangle
_{ABC}}$). Generally speaking, if one of the four-photon coincidence detectors clicks ($T$,
$e_1$, $e_2$, and $E_3$ or $T$, $E_1$, $E_2$, and $e_3$), we can, at last, obtain the desired three-photon GHZ state by postselection and a bit-flip operation on the appropriate photon.

Channel noise can affect a system due to the dissipative interaction of states with the environment and can destroy the entangled states. For both polarization entanglement and spatial entanglement, a pure state will always become a mixed one in practice. Therefore, the three photons may suffer from channel noise when traveling from a source to a destination (i.e., $d_i$ and $D_i$ ($i=1,2,3$)). To simplify the process, we assume that the photons only suffer depolarization consisting of both bit-flip and phase errors, as described by Simon and Pan in Ref.\cite{SimonPan2002}. Thus, in a noisy channel, the original state $\left| {{\psi ^ + }} \right\rangle _{PS}$ may be transformed to one of the following eight three-photon quantum states:
\begin{equation}\label{}
\left\{ \begin{gathered}
  {\left| {\psi _{}^ \pm } \right\rangle _{PS}} = \frac{1}
{2}\left( {\left| {HHV} \right\rangle \left( {{d_1}{d_2}{D_3} + {D_1}{D_2}{d_3}} \right) \pm \left| {VVH} \right\rangle \left( {{d_1}{D_2}{d_3} + {D_1}{d_2}{D_3}} \right)} \right) \hfill, \\
  {\left| {\psi _0^ \pm } \right\rangle _{PS}} = \frac{1}
{2}\left( {\left| {HHH} \right\rangle \left( {{d_1}{d_2}{D_3} + {D_1}{D_2}{d_3}} \right) \pm \left| {VVV} \right\rangle \left( {{d_1}{D_2}{d_3} + {D_1}{d_2}{D_3}} \right)} \right) \hfill, \\
  {\left| {\psi _1^ \pm } \right\rangle _{PS}} = \frac{1}
{2}\left( {\left| {VHH} \right\rangle \left( {{d_1}{d_2}{D_3} + {D_1}{D_2}{d_3}} \right) \pm \left| {HVV} \right\rangle \left( {{d_1}{D_2}{d_3} + {D_1}{d_2}{D_3}} \right)} \right) \hfill, \\
  {\left| {\psi _2^ \pm } \right\rangle _{PS}} = \frac{1}
{2}\left( {\left| {HVH} \right\rangle \left( {{d_1}{d_2}{D_3} + {D_1}{D_2}{d_3}} \right) \pm \left| {VHV} \right\rangle \left( {{d_1}{D_2}{d_3} + {D_1}{d_2}{D_3}} \right)} \right) \hfill. \\
\end{gathered}  \right.
\end{equation}
The combined system then evolves as
\begin{equation}\label{}
\left\{ \begin{gathered}
  {\left| {{\psi ^ \pm }} \right\rangle _{PS}} \to \frac{1}
{2}\left( {\left( {\left| {HHH} \right\rangle  + \left| {VVV} \right\rangle } \right)\left| {{e_1}{e_2}{E_3}} \right\rangle  \pm \left( {\left| {HVV} \right\rangle  + \left| {VHH} \right\rangle } \right)\left| {{E_1}{E_2}{e_3}} \right\rangle } \right) \hfill, \\
  {\left| {\psi _0^ \pm } \right\rangle _{PS}}\to \frac{1}
{2}\left( {\left( {\left| {HHV} \right\rangle  + \left| {VVH} \right\rangle } \right)\left| {{e_1}{e_2}{e_3}} \right\rangle  \pm \left( {\left| {VHV} \right\rangle  + \left| {HVH} \right\rangle } \right)\left| {{E_1}{E_2}{E_3}} \right\rangle } \right) \hfill, \\
  {\left| {\psi _1^ \pm } \right\rangle _{PS}} \to \frac{1}
{2}\left( {\left( {\left| {VHV} \right\rangle  + \left| {HVH} \right\rangle } \right)\left| {{E_1}{e_2}{e_3}} \right\rangle  \pm \left( {\left| {HHV} \right\rangle  + \left| {VVH} \right\rangle } \right)\left| {{e_1}{E_2}{E_3}} \right\rangle } \right) \hfill, \\
  {\left| {\psi _2^ \pm } \right\rangle _{PS}} \to \frac{1}
{2}\left( {\left( {\left| {HVV} \right\rangle  + \left| {VHH} \right\rangle } \right)\left| {{e_1}{E_2}{e_3}} \right\rangle  \pm \left( {\left| {HHH} \right\rangle  + \left| {VVV} \right\rangle } \right)\left| {{E_1}{e_2}{E_3}} \right\rangle } \right) \hfill. \\
\end{gathered}  \right.
\end{equation}
Surprisingly, in this scheme, the desired GHZ state can also be obtained by performing several appropriate unitary operations on the three photons. The outputs and the unitary operations are correspondingly presented in Table \ref{TAB 1}.

\renewcommand\arraystretch{1.5}
\begin{table}[htbp]
 \caption{Outputs and appropriate unitary operations for the case of one photon in each spatial mode $a_1$, $b_1$, $a_2$, and $b_2$, where $\hat I$ and ${\hat \sigma _x}$ respectively represent the identity operator and bit-flip operation.}\label{TAB 1}

 \tabcolsep0.15in
 \doublerulesep2pt
 \begin{tabular}{l l l l l}
  \hline\hline
 \ inputs   & outputs   & $A$ & $B$ & $C$\\
 \hline

\ ${\left| {\psi ^ \pm } \right\rangle _{PS}}$& ${e_1}{e_2}{E_3}$& $\hat I$  &$\hat I$ &${\hat \sigma _x}$\\
\ ${\left| {\psi ^ \pm } \right\rangle _{PS}}$& ${E_1}{E_2}{e_3}$& $\hat I$  & ${\hat \sigma _x}$&$\hat I$\\

\ ${\left| {\psi _0^ \pm } \right\rangle _{PS}}$& ${e_1}{e_2}{e_3}$& $\hat I$  & $\hat I$&${\hat I}$\\
\ ${\left| {\psi _0^ \pm } \right\rangle _{PS}}$& ${E_1}{E_2}{E_3}$& ${\hat \sigma _x}$  & $\hat I$&$\hat I$\\

\ ${\left| {\psi _1^ \pm } \right\rangle _{PS}}$& ${E_1}{e_2}{e_3}$& ${\hat \sigma _x}$&$\hat I$  & $\hat I$\\
\ ${\left| {\psi _1^ \pm } \right\rangle _{PS}}$& ${e_1}{E_2}{E_3}$& $\hat I$  & $\hat I$&$\hat I$\\

\ ${\left| {\psi _2^ \pm } \right\rangle _{PS}}$& ${e_1}{E_2}{e_3}$& $\hat I$  & ${\hat \sigma _x}$&$\hat I$\\
\ ${\left| {\psi _2^ \pm } \right\rangle _{PS}}$& ${E_1}{e_2}{E_3}$& $\hat I$  & $\hat I$&${\hat \sigma _x}$\\

  \hline
  \hline
 \end{tabular}

\end{table}

For $i=j$, i.e., the case of two photons each in the upper
modes $a_1$ and $b_1$ or each in the lower modes $a_2$ and $b_2$, the initial state can be written as
\begin{equation}\label{}
{\left| {{\phi ^ + }} \right\rangle _{PS}} = \frac{1} {2}\left(
{{{\left| {HHV} \right\rangle }_{{D_1}{D_2}{D_3}}} + {{\left| {HHV}
\right\rangle }_{{d_1}{d_2}{d_3}}} + {{\left| {VVH} \right\rangle
}_{{D_1}{D_2}{D_3}}} + {{\left| {VVH} \right\rangle
}_{{d_1}{d_2}{d_3}}}} \right).
\end{equation}
Under the condition that the trigger photon is ignored, the above state created by our device shown in Fig.1 is exactly equivalent to the original state shown in the Eq.(3) of Deng's paper \cite{dengfuguo2011} with $N=3$. The discussion for this case is analogous to that of Deng's in Ref.\cite{dengfuguo2011}, so we will not repeat the discussion here. Finally, we can successfully obtain the desired three-photon GHZ state (\ref{GHZ state}) from the states (\ref{up down}) and (\ref{four-mode}) which are entangled in the spatial and polarization degrees of freedom.

\section{summary}

In this paper we have explored an efficient scheme for creating three-photon Greenberger-Horne-Zeilinger state with currently available linear optical elements and parametric down-conversion sources. In previous scheme \cite{GHZExperiment99} where three-photon GHZ entanglement have been obtained experimentally, two pairs of entangled photons were used to create three-photon GHZ state and came from the upper modes $a_1$ and $b_1$ (i.e. case 1 of the present scheme). In our scheme, two pairs are collected for both the upper modes $a_1$ and $b_1$ and the lower modes $a_2$ and $b_2$ as well as the four-mode $a_1$, $b_1$, $a_2$, and $b_2$. Thus, two kinds of three-photon GHZ states in both polarization and spatial entanglement can be created in principle.

In order to generate the desired GHZ state, we improved the scheme with several half wave plates and PBSs. The heart of the scheme is the device of {\sc GHZps} which is used to create available entangled states. The core element of {\sc GHZps} is a two-photon QND detector which is used to determine the four photons created by two PDC sources whether in the state (\ref{up down}) or in the alternative state (\ref{four-mode}). Focusing particularly on the state (\ref{four-mode}), it has entanglement in both the polarization and the spatial degrees of freedom, but it is not a product state in the two degrees of freedom. Finally, a large number of the desired three-photon GHZ states can be obtained by postselection and several appropriate unitary operations. This process works not only in a noiseless quantum channel but also in noisy channel (depolarization) situations. Therefore, this paper proposes a device that should have two quite remarkable advances: one is a higher yield rate and another is a powerful error-correcting capability in noisy channels.

This work was supported by the National Natural Science Foundation
of China under Grant No: 10971247, Hebei Natural Science Foundation
of China under Grant Nos: F2009000311, A2010000344, the Fundamental Research Funds for the Central Universities of China under Grant No:2011B025.

\end{document}